%% file: lp208-nuphys2015-proceedings.tex
\documentclass[12pt]{article}
\usepackage{graphicx}
\usepackage{xcolor}
\usepackage{amssymb}
\usepackage{amsmath}
\usepackage{nicefrac}

\newcommand\pubnumber{SNSN-323-63}
\newcommand\pubdate{\today}

\def\imperialpd{Imperial College London, Department of Physics,
South Kensington Campus, London SW7 2AZ}
\def\oxfordpd{Department of Physics, Oxford University, Oxford, Oxfordshire,
United Kingdom}
\def\Title#1{\begin{center} {\Large #1 } \end{center}}
\def\Author#1{\begin{center}{ \sc #1} \end{center}}
\def\Address#1{\begin{center}{ \it #1} \end{center}}

\newcommand\pubblock{\rightline{\begin{tabular}{l} \pubnumber\\
         \pubdate  \end{tabular}}}
\newenvironment{Abstract}{\begin{quotation}  }{\end{quotation}}
\newenvironment{Presented}{\begin{quotation} \begin{center}
             PRESENTED AT\end{center}\bigskip
      \begin{center}\begin{large}}{\end{large}\end{center} \end{quotation}}
\def\Acknowledgements{\bigskip \bigskip \begin{center} \begin{large}
             \bf ACKNOWLEDGEMENTS \end{large}\end{center}}

\input econfmacros.tex

\begin{document}
\begin{titlepage}
\pubblock

\vfill
\Title{Theoretical predictions of transverse kinematic imbalance in
neutrino-nucleus interactions}
\vfill
\Author{Luke Pickering}
\Address{\imperialpd}
\Author{Xianguo Lu}
\Address{\oxfordpd}
\vfill
\begin{Abstract}
Distributions of transverse kinematic imbalance in neutrino-nucleus
interactions in the few GeV regime are sensitive to nuclear effects.
We present a study comparing the latest predictions of
transverse kinematic imbalance from the
interaction simulations, NuWro and GENIE.
We discuss the differences between the model predictions.
\end{Abstract}
\vfill
\begin{Presented}
NuPhys2015, Prospects in Neutrino Physics\\
Barbican Centre, London, UK,  December 16--18, 2015
\end{Presented}
\vfill
\end{titlepage}
\def\thefootnote{\fnsymbol{footnote}}
\setcounter{footnote}{0}

\newcommand{\dpt}{{\delta\it{p_\mathrm{T}}}}
\newcommand{\dat}{{\delta\it{\alpha_\mathrm{T}}}}
\newcommand{\dphit}{{\delta\it{\phi_\mathrm{T}}}}
\newcommand{\mupt}{{\it{p}^\ell_\mathrm{T}}}
\newcommand{\dptt}{{\delta\it{p_\mathrm{TT}}}}
\newcommand{\tdpt}{$\dpt$}
\newcommand{\tdat}{$\dat$}
\newcommand{\tdphit}{$\dphit$}
\newcommand{\tmupt}{$\mupt$}
\newcommand{\tdptt}{$\dptt$}

\newcommand{\qsq}{Q^{2}}
\newcommand{\tqsq}{$\qsq$}

\newcommand{\enu}{E_{\nu}}
\newcommand{\tenu}{$\enu$}

\newcommand{\etrans}{\omega}
\newcommand{\tetrans}{$\etrans$}

\newcommand{\hadrmass}{W}
\newcommand{\thadrmass}{$\hadrmass$}

\newcommand{\figtxt}{Fig.~}

\newcommand{\needcite}[1]{ {\color{red}$[#1]$} }

\subsubsection*{Introduction}

\begin{figure}[htb]
\centering
\includegraphics[height=1.75in]{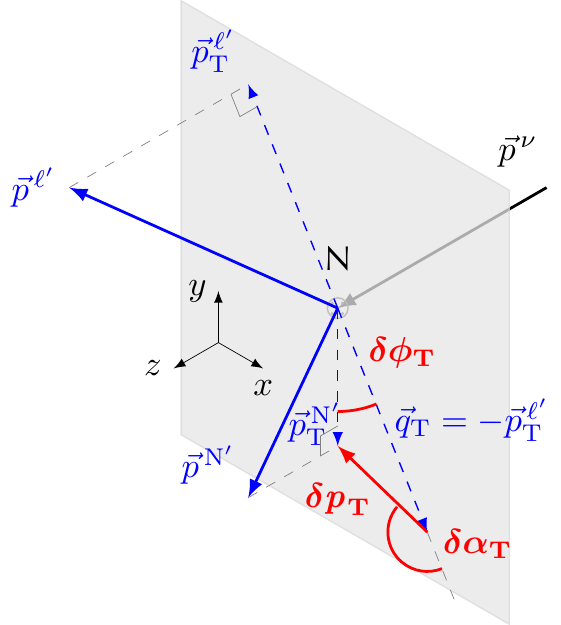}
\hspace{4em}
\includegraphics[height=1.75in]{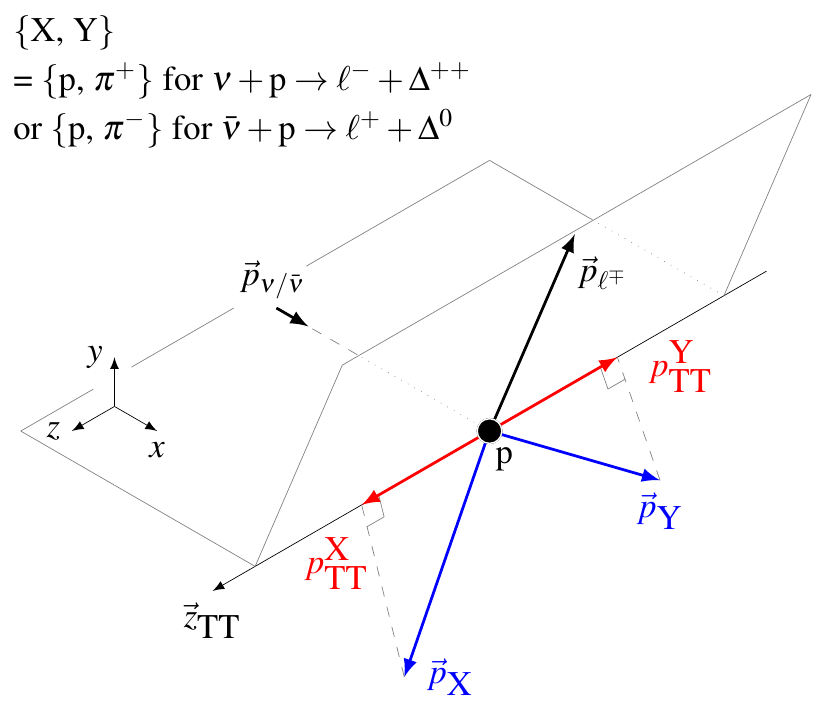}
\caption{Single- (left)
    and double- (right) transverse kinematics. The variables \tdpt, \tdat, \tdphit, and $\dptt \equiv \it{p}_\textrm{TT}^\textrm{Y} + \it{p}_\textrm{TT}^\textrm{Y}$ each
    represents a departure from the kinematics of elementary neutrino interactions
    on stationary, free nucleons. Figures taken from \cite{tttpaper,Lu:2015vri}.}
\label{fig:tvdef}
\end{figure}

Neutrino interaction models currently constitute a significant proportion of
future few GeV neutrino experimental uncertainty budget.
Conventionally, investigation has focused on how using nuclear targets
affects the charged lepton kinematics---as opposed to interactions on free
nucleon targets.
Such effects can be conflated with unknown neutrino energy
for neutrino beams produced by accelerators.
Single-transverse kinematic imbalances exhibit a significantly reduced
dependence on neutrino energy and data measurements will provide new insight
into a number of nuclear effects \cite{tttpaper,nuint15proc}.
The double-transverse kinematic imbalance,  \tdptt\,   provides a novel method of
reconstructing neutrino energy spectra independent of nuclear effects
\cite{hydrogenpaper,Lu:2015vri}. The definitions of the single-transverse
variables \tdpt, \tdphit, \tdat, as well as \tdptt\ are shown in
Figure~\ref{fig:tvdef}.

The following predictions are generated using GENIE 2.10.0 \cite{GENIE} with the
nominal hA FSI model, and NuWro 11q \cite{NuWro}. The predicted distributions are
generated using the NuMI on-axis $\nu_\mu$ and $\bar{\nu}_\mu$ flux shapes.

\subsubsection*{Single transverse kinematic imbalance in neutrino quasi-elastic scattering}

\begin{figure}[htb]
\centering
\includegraphics[height=1.5in]{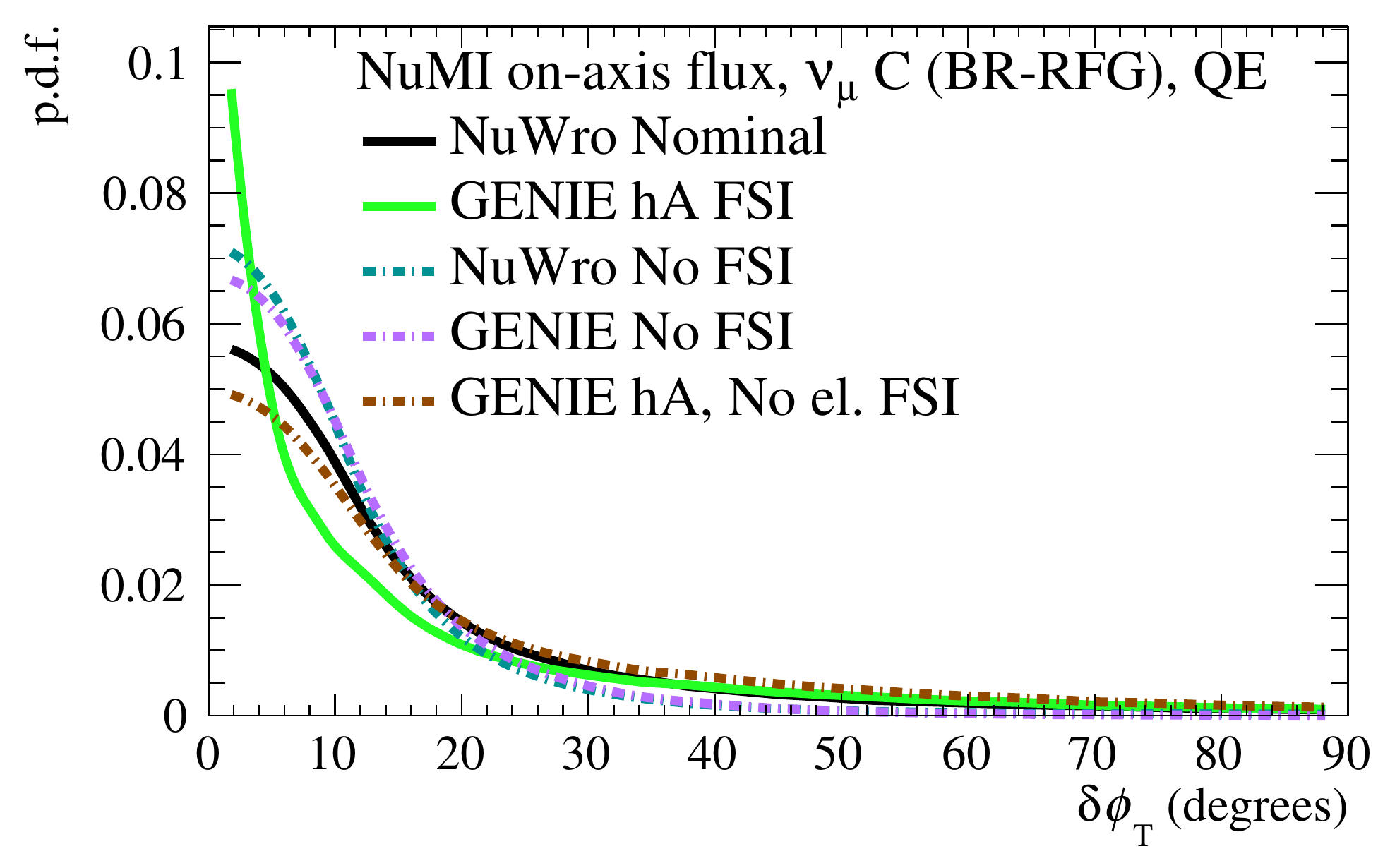}
\includegraphics[height=1.5in]{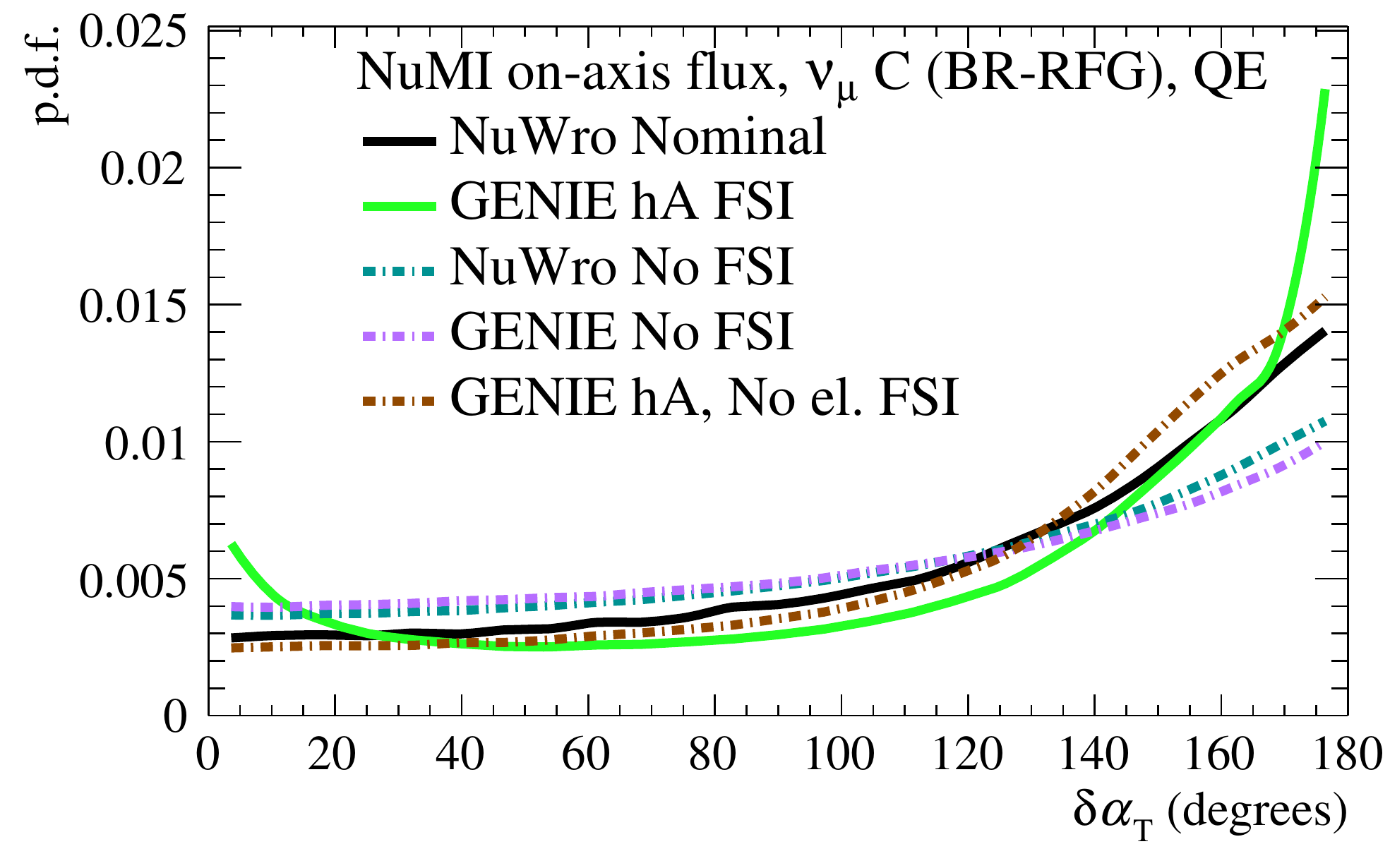}
\caption{The NuWro and GENIE predictions for \tdphit\ (left) and
    \tdat\ (right). Nominal distributions are compared to the cases where FSI is disabled. Further comparison is made by removing nominal GENIE events that experienced proton elastic FSI (see text for exact definition). }
\label{fig:stvnuqe}
\end{figure}

The observable \tdphit\ characterises how `back-to-back' the
transverse components of the final states are.
In the absence of FSI and multi-nucleon correlations, the only source of
 transverse kinematic imbalance should be the transverse component of
the Fermi motion of the struck nucleon.
This distribution has been measured in neutrino scattering
before~\cite{nomaddphit, minervadphit, t2kdphit}, most recently by the MINER$\nu$A collaboration
which presents a measurement of $\varphi = 180^{\circ} - \dphit$ compared to a
GENIE simulation. MINER$\nu$A found a good agreement between the data and the GENIE simulation\footnote{The version of GENIE used, 2.6.2,
did not contain an elastic FSI component}.
The predictions from NuWro and GENIE are shown in Figure~\ref{fig:stvnuqe}.
The left panel shows that the nominal GENIE simulation
predicts a very sharp back-to-back peak. The most striking
feature is that the enhancement with respect to the NuWro prediction around $\dphit=0$ is not evident in the GENIE
`No FSI' curve. The full simulation appears to induce less transverse imbalance
than the case with hadronic re-interactions  disabled.

The `transverse boosting angle', \tdat, shows the apparent `acceleration' or
`deceleration' of the hadronic final state arising from nuclear effects.
For $\dat > 90^\circ$, \tdpt\ points in a similar direction as the charged
lepton---the hadronic final state has less transverse momentum than is
expected from the free nucleon target case.
The effect of intra-nuclear re-interactions is expected to be an energy-momentum
transfer to the nuclear medium---a deceleration of the interacting hadronic
state---which corresponds to a peak at $\dat \sim 180^\circ$.
While both NuWro and GENIE predict this behaviour, as shown in
Figure~\ref{fig:stvnuqe} (right), the GENIE prediction exhibits a
significantly sharper deceleration peak and a less prominent peak at $\dat \sim 0$
that corresponds to some accelerating effect.
\begin{figure}[htb]
\centering
\includegraphics[height=1.5in]{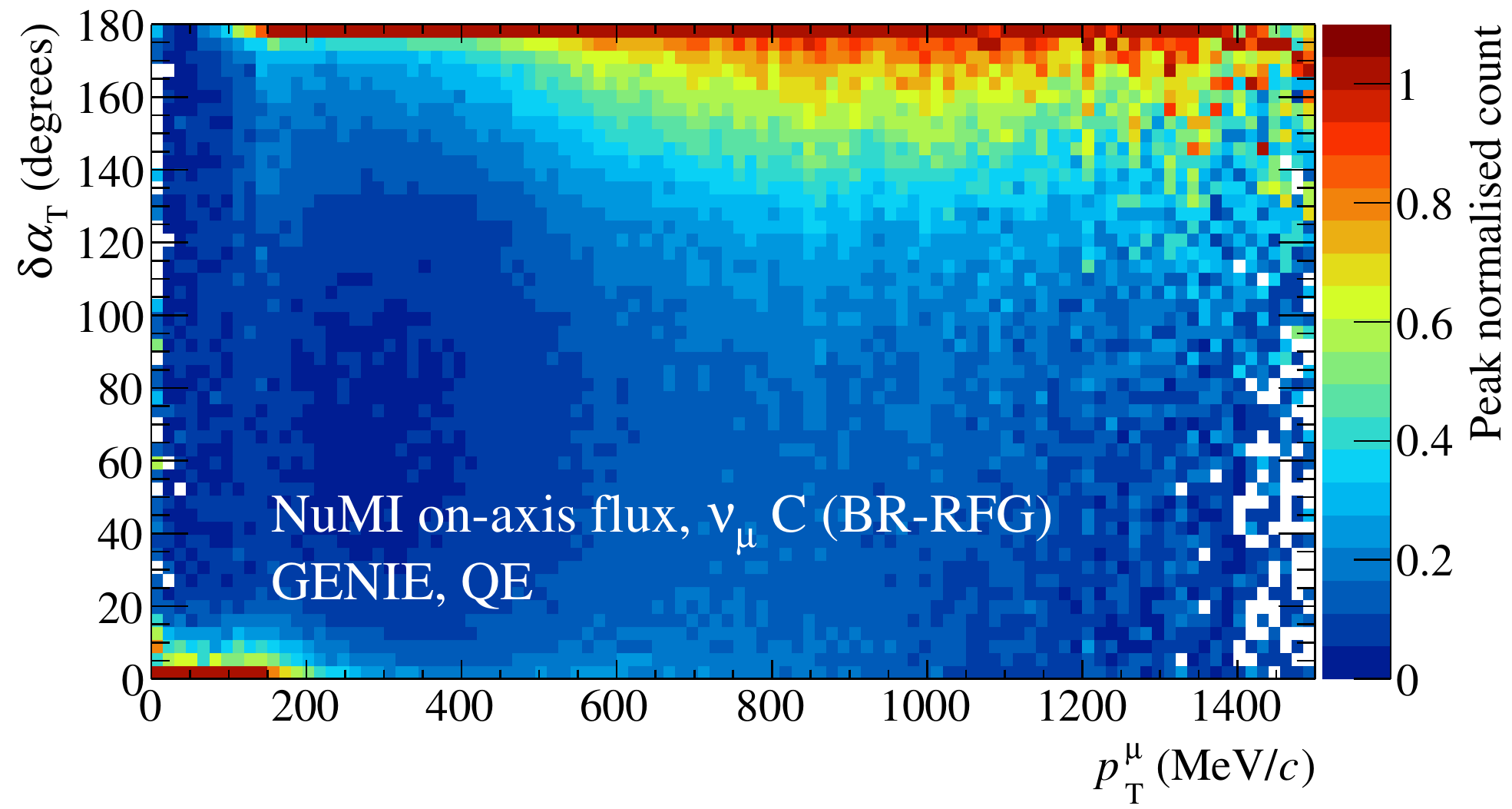}
\includegraphics[height=1.5in]{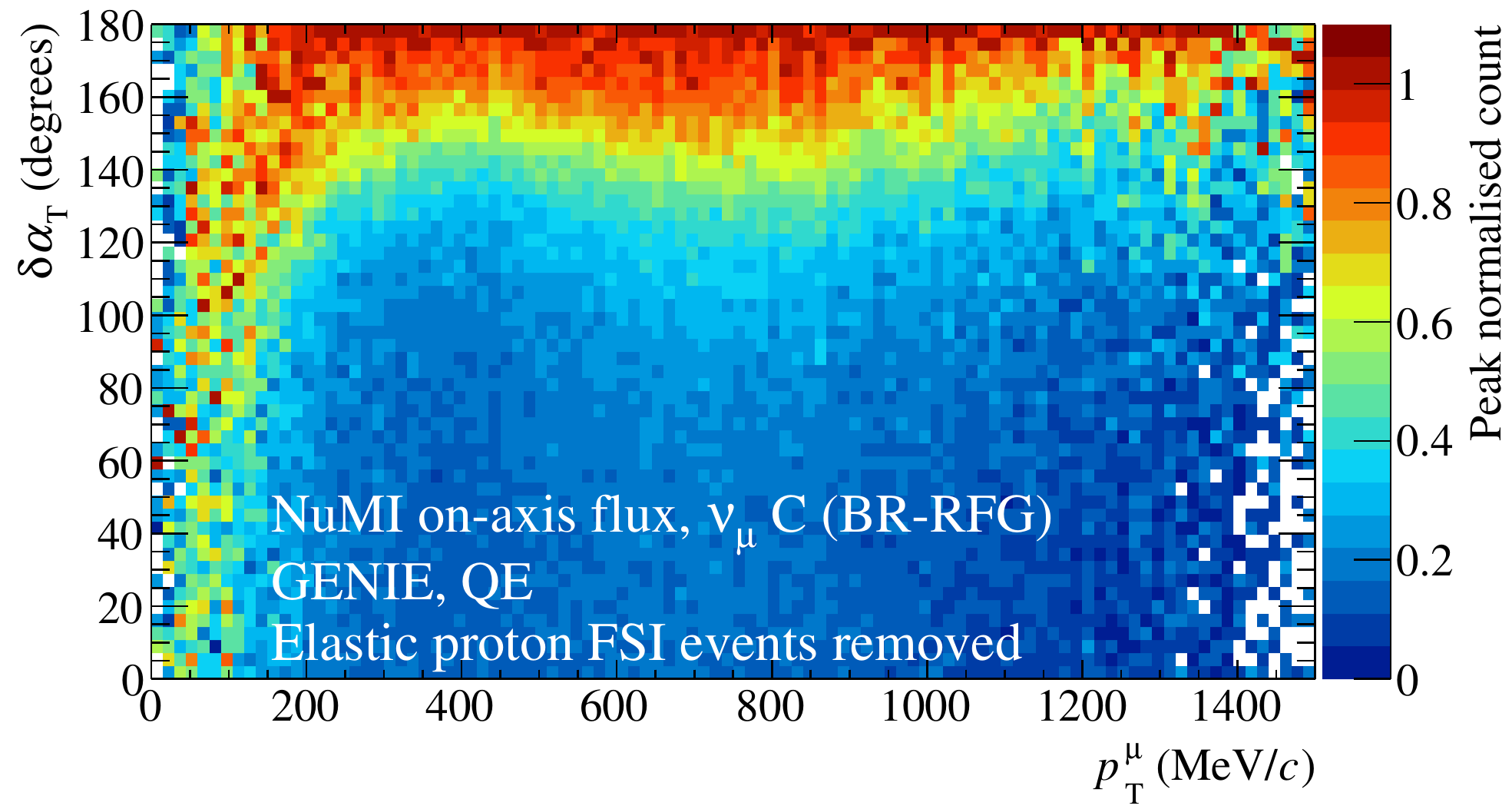}
\caption{The variation of \tdat\ with \tmupt\ for QE events, as predicted by
GENIE, with (left) and without (right) elastic proton FSI.}
\label{fig:geniedatmupt}
\end{figure}
 It is useful to investigate transverse imbalance as a function of
the charged lepton transverse momentum\footnote{This separates extra
neutrino energy dependence caused by the \tmupt\ dependence of \tdphit\
\cite{tttpaper,nuint15proc}}.
Figure~\ref{fig:geniedatmupt} shows the GENIE \tdat\ distribution in
slices of lepton transverse momentum. Each \tmupt\
slice is normalised such that the most probable value is set to unity.
The left panel shows that the accelerating peak in the \tdat\
prediction is only evident at low \tmupt.

As suggested by the GENIE collaboration, we investigated removing events which
include a final state proton that underwent an elastic interaction defined in the hA FSI model.
The nominal GENIE simulation predicts that such events amount to about
$40\%$ of QE interactions at the NuMI beam energy.
Having removed such events, both the \tdphit\ and
\tdat\ distributions are more similar to the NuWro prediction (Figure~\ref{fig:stvnuqe}).
The sharp peaks in \tdphit\ and \tdat\ are notably reduced.
The right panel of Figure~\ref{fig:geniedatmupt} shows \tdat\ as a
function of \tmupt\ with elastic FSI events removed. The GENIE hA elastic FSI
model causes proton final state acceleration for low \tmupt\ and strong
deceleration for $\mupt \gtrsim 200\,\textrm{MeV}/c$. The resulting sharp peaks
at $\dphit = 0$ and $\dat = $0, 180 degrees indicate a ($q_\textrm{T}$-dependent\footnote{See Figure~\ref{fig:tvdef} (left) for definition of $q_\textrm{T}$.}) strong collinear enhancement in the proton intra-nuclear scattering  cross section.


\subsubsection*{Transverse kinematic imbalance in neutrino-induced resonance production}

\begin{figure}[htb]
\centering
\includegraphics[height=1.75in]{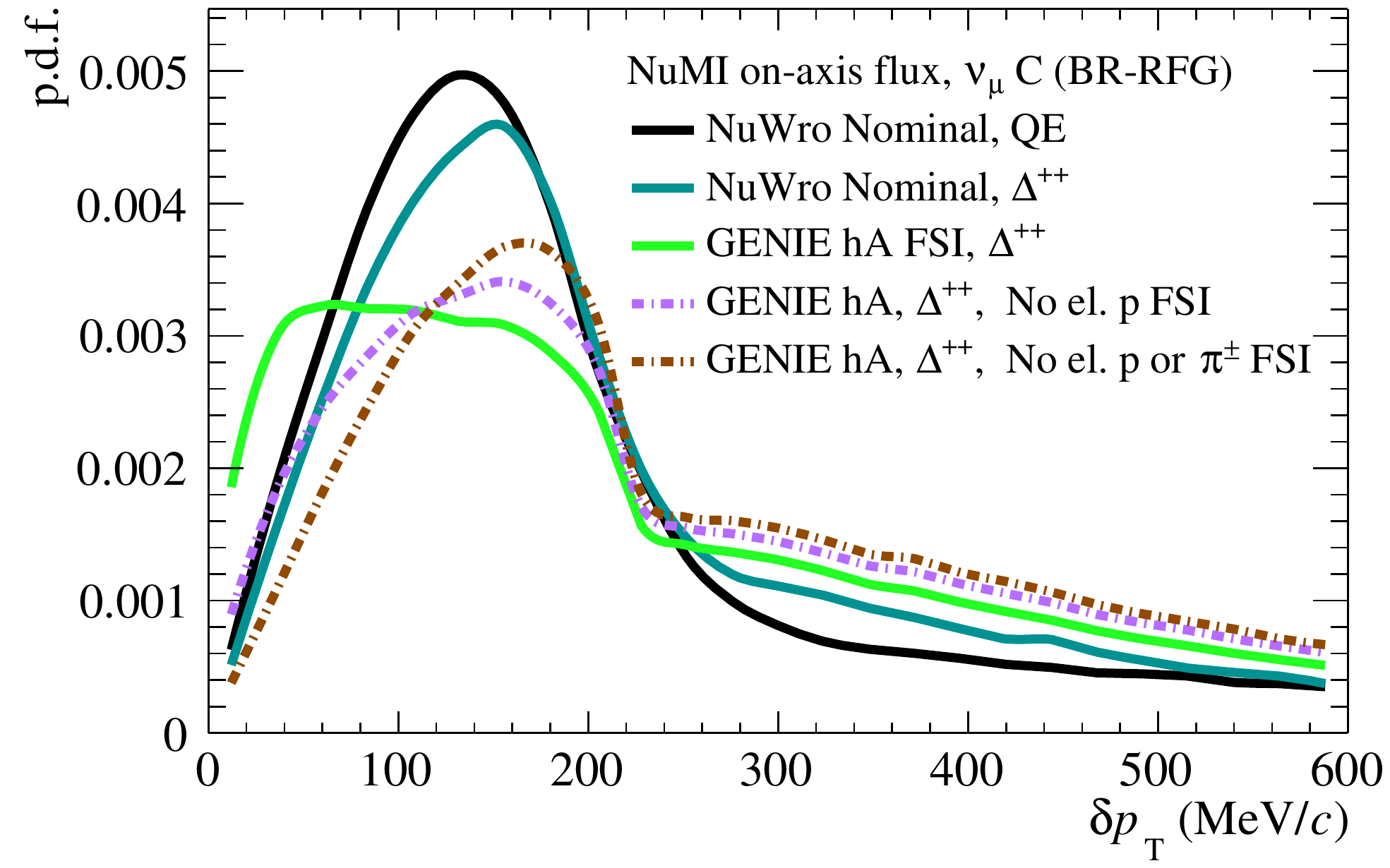}
\includegraphics[height=1.75in]{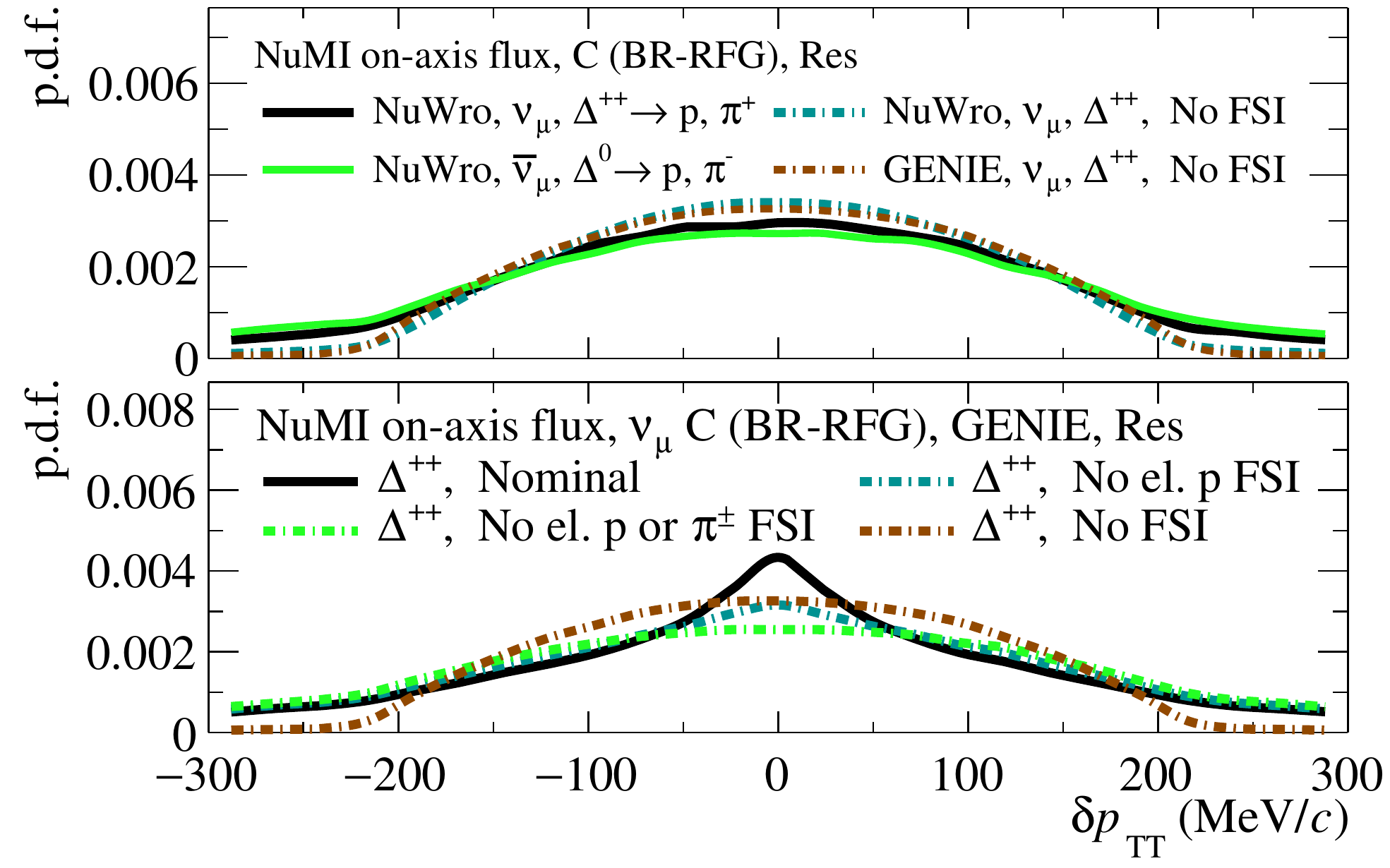}
\caption{The NuWro and GENIE predictions for \tdpt\ (left) and
    \tdptt\ (right). The features shown in the lower right panel are also exhibited in the p$\pi^-$ and p$\pi^0$ channels.}
\label{fig:resstv}
\end{figure}

In resonance production, where the intermediate resonant state decays within
the nucleus to multiple hadrons, nuclear effects change the kinematics of all
hadronic final states.
Details of such effects can be studied in p$\pi$ channels as follows and provide new insight that is not accessible in QE interactions.

With reference to Figure~\ref{fig:tvdef}, $\vec{p}^{\,\textrm{N}^\prime}$ becomes
$\vec{p}^{\,\textrm{RES}} = \vec{p}^{\,\textrm{p}} + \vec{p}^{\,\pi}$.
This opens up three new event selections,
$\nu_{\ell} + \textrm{p} \xrightarrow{\Delta^{++}} \textrm{p} + \pi^+ + \ell^-$,
$\nu_\ell + \textrm{n} \xrightarrow{\Delta^{+}} \textrm{p} + \pi^{0} + \ell^-$,
and
$\bar{\nu}_{\ell} + \textrm{p} \xrightarrow{\Delta^{0}} \textrm{p} + \pi^- + \ell^+$.
The $\dpt$ predictions for QE and $\Delta^{++}$ production are shown in
Figure~\ref{fig:resstv} (left).
The shape difference evident in the GENIE prediction shows that the effects of
the elastic FSI component are not confined to nucleon FSIs but also exist for
pions as well. Because the  same model (BR-RFG~\cite{GENIE, NuWro}) is used for the nuclear state in all simulations, a stronger FSI in GENIE can be inferred by its higher proportion of events with $\dpt \gtrsim 250\,\textrm{MeV}/c$.

Figure~\ref{fig:resstv} (right) shows the GENIE and NuWro predictions
for \tdptt. The top panel shows that
without FSI both are consistent, and that the NuWro
distribution widens as re-scattering takes place when FSI is enabled.
The bottom panel shows the effect of the GENIE elastic component. Elastic re-interactions, in both nucleon and
pion FSI of the hA model, result in an enhancement around $\dptt = 0$.


\subsubsection*{Summary and Outlook}

The GENIE and NuWro predictions for a number of transverse kinematic imbalances
have been shown. The phenomenological predictions exhibit significant shape
differences in important regions of the distributions.
Measurements of these observables are underway and should
offer separation power among models of nuclear effects.
A better understanding of
the hadronic cascade, constrained by data measurement, will result in reduced
systematic uncertainty for future
neutrino cross-section and oscillation measurements.


\Acknowledgements
We express our gratitude to the NuWro Collaboration and the GENIE Collaboration,
as well as to S. Dytman, Y. Hayato, K. McFarland and  C. Wilkinson, for helpful
discussions and suggestions.

\end{document}

%% file: econfmacros.tex
\def\beq{\begin{equation}}
\def\eeq#1{\label{#1}\end{equation}}
\def\eeqn{\end{equation}}

\def\beqa{\begin{eqnarray}}
\def\eeqa#1{\label{#1}\end{eqnarray}}
\def\eeqan{\end{eqnarray}}

\let\bar=\overbar

\def\Dslash{\not{\hbox{\kern-4pt $D$}}}
\def\dslash{\not{\hbox{\kern-2pt $\del$}}}

\def\msb{{\bar{\ssstyle M \kern -1pt S}}}

